\documentclass[3p,times,twocolumn]{elsarticle}

\usepackage{ecrc}

\usepackage{graphicx, subfigure}

\volume{00}

\firstpage{1}

\journalname{Nuclear Physics B Proceedings Supplement}

\runauth{Igor V. Gorelov}

\jid{nuphbp}

\jnltitlelogo{Nuclear Physics B Proceedings Supplement}

\usepackage{amssymb}
\usepackage{eqnarray,amsmath}

\usepackage{xspace} % To avoid problems with missing or double spaces after

\usepackage[figuresright]{rotating}

\input{symblibrary.tex}

\begin{document}

\begin{frontmatter}

%% Title, authors and addresses

%% use the tnoteref command within \title for footnotes;
%% use the tnotetext command for the associated footnote;
%% use the fnref command within \author or \address for footnotes;
%% use the fntext command for the associated footnote;
%% use the corref command within \author for corresponding author footnotes;
%% use the cortext command for the associated footnote;
%% use the ead command for the email address,
%% and the form \ead[url] for the home page:
%%
%% \title{Title\tnoteref{label1}}
%% \tnotetext[label1]{}
%% \author{Name\corref{cor1}\fnref{label2}}
%% \ead{email address}
%% \ead[url]{home page}
%% \fntext[label2]{}
%% \cortext[cor1]{}
%% \address{Address\fnref{label3}}
%% \fntext[label3]{}

\dochead{}
%% Use \dochead if there is an article header, e.g. \dochead{Short communication}

\title{Heavy Flavor Physics at ATLAS and CMS}

%% use optional labels to link authors explicitly to addresses:
%% \author[label1,label2]{<author name>}
%% \address[label1]{<address>}
%% \address[label2]{<address>}

\author[unm]{Igor V. Gorelov\fnref{foratlcms}}
\ead{Igor.Gorelov@cern.ch}

\fntext[foratlcms]{This talk is presented on behalf of the ATLAS and
  CMS~Collaborations,~CERN.}

\address[unm]{Department of Physics and Astronomy, University of New Mexico, Albuquerque, NM 87131, USA}

\begin{abstract}
  Recent results on heavy flavor physics using data from the ATLAS and CMS
  detectors are presented. The searches for new physics signatures in
  \(CP\) violation of \(\BsaBs \) mixing and in \(\Bd\ra\Kstarz\mumu \)
  decays are discussed. The bottomonium and open-\b production results
  obtained from \(\propro \) collisions at LHC are shown. The results
  are based on data samples containing \(\mumu \) pairs collected with
  the ATLAS or CMS detectors by their corresponding muon trigger systems.
\end{abstract}

\begin{keyword}
%% keywords here, in the form: keyword \sep keyword
bottom quark \sep rare decays \sep CP-violation \sep heavy flavor production
%% MSC codes here, in the form: \MSC code \sep code
%% or \MSC[2008] code \sep code (2000 is the default)
\PACS 14.65.Fy \sep 12.15.Ff \sep 13.20.He \sep 13.25.Hw \sep 12.38.Qk

\end{keyword}

\end{frontmatter}

%%
%% Start line numbering here if you want
%%
% \linenumbers

%% main text
%\section{}
%\label{}

%% The Appendices part is started with the command \appendix;
%% appendix sections are then done as normal sections
%% \appendix

%% \section{}
%% \label{}

%% References
%%
%% Following citation commands can be used in the body text:
%% Usage of \cite is as follows:
%%   \cite{key}         ==>>  [#]
%%   \cite[chap. 2]{key} ==>> [#, chap. 2]
%%
%
\section{\(CP\)-violating weak phase \(\phis\) and \(\DGs\) from flavor-tagged
         time-dependent angular analysis of \(\Bst\) by ATLAS}
\label{delgamma}
  New phenomena beyond the predictions of the Standard Model (SM) may
  alter \(CP\) violation in \(B\)-decays. A channel that is expected to
  be sensitive to new physics contributions is the decay
  \(\overline{\Bs}\to\Jpsi\phi\) (quark content
  \(\aBs\equiv\b\bar{\s}\)~\cite{Beringer:1900zz}).
  \(CP\) violation in the \(\overline{\Bs}\to\Jpsi\phi\) decay
  occurs due to interference between direct decays and decays occurring
  through \BsaBs mixing.  The oscillation frequency of \Bs\ meson mixing
  is characterized by the mass difference \dms\ of the heavy (\BH) and
  light (\BL) mass eigenstates. The \(CP\)-violating phase \phis\ is
  defined as the weak phase difference between the \( \BsaBs\) mixing
  amplitude and the \(b\to{c}\cbar{s}\) decay amplitude. In the SM the
  phase \phis\ is small and can be related to CKM quark mixing matrix
  elements; a value of
  \(\phis\simeq-2\beta_{s}=-0.0368\pm0.0018\)~rad
  is predicted in the SM~\cite{Bona:2006sa}.
  Many models describing a new physics predict
  larger \(\phis\) values whilst satisfying all existing constraints,
  including the precisely measured value of
  \dms\ \cite{Abulencia:2006ze, Aaij:2011qx}.  Another physical quantity
  involved in \(\BsaBs\) mixing is the width difference
  \(\DGs=\GL-\GH\) of \(\BL\) and \(\BH\).  Physics beyond the SM is not
  expected to affect \DGs\ as significantly as \phis~\cite{Lenz:2011ti}.
  The decay of the pseudoscalar \(\aBs\) to the vector-vector
  final-state \(J/\psi\phi\) results in an admixture of \(CP\)-odd and
  \(CP\)-even states, with orbital angular momentum \(L = 0\), \(1\) or
   \(2\). The \(CP\) states are separated statistically through the
  time-dependence of the decay and angular correlations among the
  final-state particles.
\par
  The analysis is based on a data sample of an integrated luminosity
  \(4.9\,\ifb \) collected in 2011 by the ATLAS detector in \(\propro\)
  collisions at \(\sqrt{s} = 7\TeV\) with di-muon triggers selecting
  \(\Jpsi\to\mumu\) candidates~\cite{atlas-conf-phis}.  The triggers
  select di-muon events requiring both muons to have
  \(\pt(\mupm)>4\gevc\), or with asymmetric requirements of
  \(\pt(\mu_{1})>6\gevc\), and \(\pt(\mu_{1})>4\gevc\) with a rapidity
  range of \(\left|\eta(\mupm)\right|<2.4\) for both cases.  The
  candidates for \(\phi\ra\Km\Kp \) are reconstructed from all pairs of
  oppositely charged particles with \(\pt(h)>1\gevc\) and
  \(\left|\eta(h)\right|<2.5\), that are not identified as muons.
  The \(\aBs\) candidates are reconstructed using measurements provided
  by the inner tracking detectors and the muon
  spectrometers~\cite{Aad:2008zzm}.
  Candidates for
  \(\aBs\ra\Jpsi(\ra\mumu)\phi(\ra\Km\Kp)\)~\footnote{Unless otherwise
    stated all references to the specific charge combination imply the
    charge conjugate combination as well.}
  are sought by fitting the tracks for each combination of
  \(\Jpsi\to\mumu \) and \(\phi\to\Km\Kp\) to a common vertex with
  \(m(\mumu)\) constrained to the world average 
  \(m(\Jpsi)\)~\cite{Beringer:1900zz}.
  In total \(N(\Bs)=131,000\) candidates are collected within a mass
  range of \(m(\Bs)\in(5.15,\,5.65)\gevcc\) to be used in the likelihood
  fit.
  To infer the initial flavor of the \(\B\)-meson (\(\aBs\) or \(\Bs\))
  the flavor of the \(\B\)-hadron originating from the other \(b\)-quark
  (OS tagging) is determined.  The flavor tagging probabilities are
  evaluated using the calibration mode \(\Bub\to{J}/{\psi{K}^{-}}\)
  reconstructed in the same data sample.
  A maximum likelihood fit is performed over an unbinned set of the
  reconstructed mass \(m\), the measured proper decay time, 
  \(t\equiv\lxy\cdot{m}(\Bs)/({c}\cdot\pt(\Bs))\), the
  measured mass and proper decay time uncertainties \(\sigma_{m}\) and
  \(\sigma_{t}\), and the transversity angles \(\Omega\) of each 
  \(\overline{\Bs}\to\Jpsi\phi\) decay
  candidate satisfying the selection criteria~\cite{atlas-conf-phis}.
  The fit finds the following values for the physics parameters of the interest:
%
%\begin{eqnarray}
\begin{align*}
   \phis                & = 0.12\pm0.25\,\rm{(stat.)}\pm0.11\,\rm{(syst.)}\,rad\\
   \DGs                 & = 0.053\pm0.021\,\rm{(stat.)}\pm0.009\,\rm{(syst.)}\,{ps}^{-1}\\ 
   \Gs                  & = 0.677\pm0.007\,\rm{(stat.)}\pm0.003\,\rm{(syst.)}\,{ps}^{-1}\\
   |A_{0}(0)|^2         & = 0.529\pm0.006\,\rm{(stat.)}\pm0.011\,\rm{(syst.)}\\
   |A_{\parallel}(0)|^2 & = 0.220\pm0.008\,\rm{(stat.)}\pm0.009\,\rm{(syst.)}\\
   {\delta_{\perp}}     & = 3.89\pm0.46\,\rm{(stat.)}\pm0.13\,\rm{(syst.)}\,{rad}
\end{align*}
%\end{eqnarray}
% 
%
  The resulting contours for the several confidence intervals are
  produced using a profile likelihood method and are shown in
  Fig.~\ref{fig:Contour}.
  \begin{figure}[hbt]
  \begin{center}
    \includegraphics[width=0.34\textwidth]{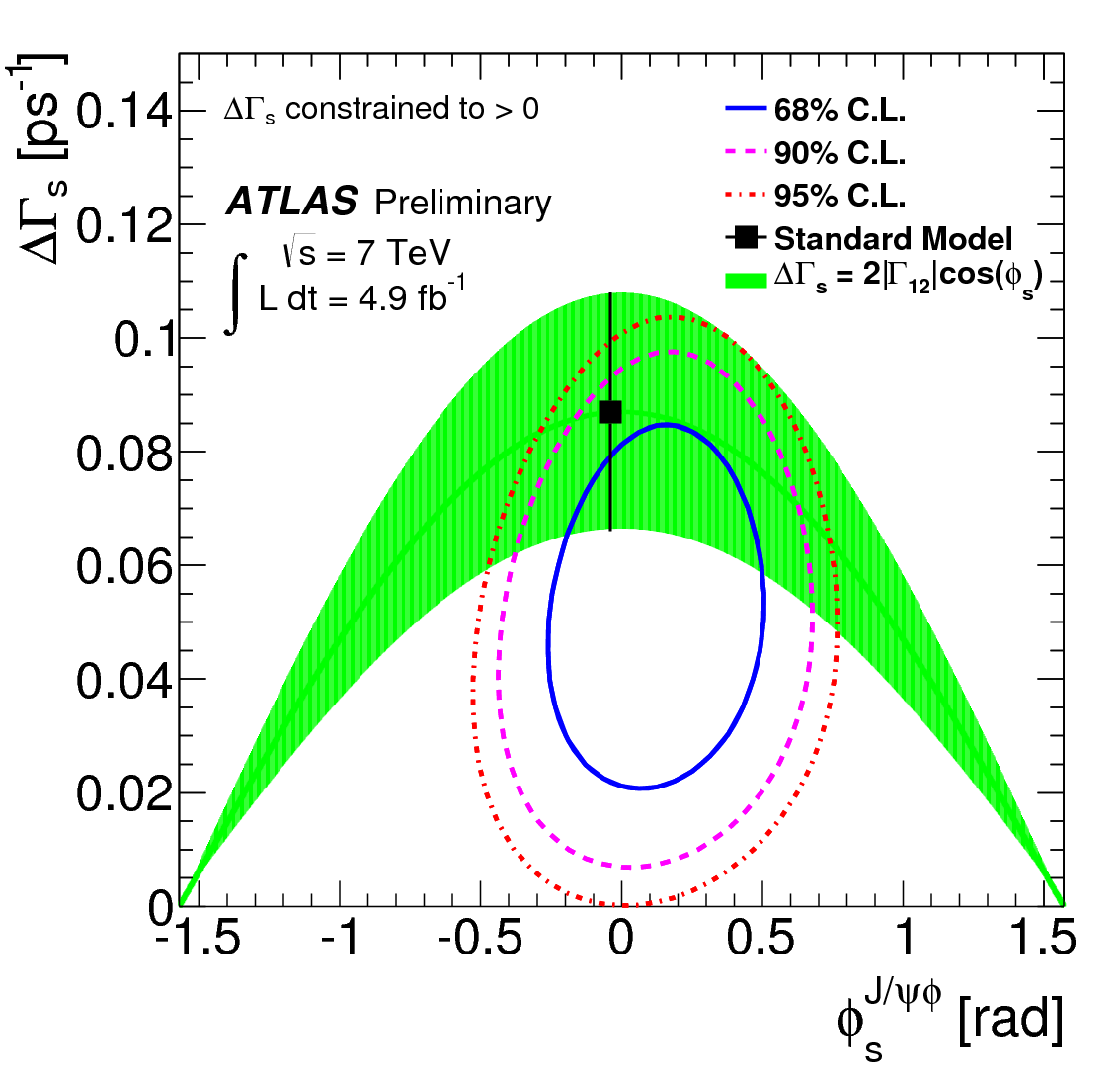}
    \caption{Likelihood contours in the \(\phis{-}\DGs\) plane.  Three
             contours show the \(68\%,\,90\%\) and \(95\%\) confidence intervals
             (statistical errors only). }
  \end{center}
  \label{fig:Contour}
  \end{figure}
  The values are consistent with those obtained in the previous untagged
  analysis~\cite{Aad:2012kba}, and as expected improving significantly
  on the overall uncertainty on \(\phis\). These results are also
  consistent with theoretical expectations, in particular \(\phis\) and
  \DGs\ are in good agreement with the
  values predicted in the Standard Model.
\section{Angular analysis of the decay \(\Bd\ra{K^{*0}}\mumu\)}
\label{kstarmumu}
   Another productive area for indirect searches of new phenomena, in
   flavor physics, is the study of flavor-changing neutral current
   decays of \(b \) hadrons such as the semileptonic decay mode
   \(\Bdb\ra{\Kstarzb}\mumu\).
   This decay is forbidden at tree level in the SM, resulting in small
   SM rates. From the theoretical side, robust calculations are now
   possible for much of the phase space of this decay and the
   calculations also indicate that new physics could give rise to
   readily observable effects. Finally, this decay mode is relatively
   easy to select and reconstruct at hadron colliders.  Two important
   observables
   in the \(\Bdb\ra{\Kstarzb{(892)}}(\ra{\Km\pip})\mumu\) 
   decay are the forward-backward asymmetry of the muons, \({A_{FB}}\),
   and the longitudinal polarization fraction of the
   \(\Kstarzb{(892)}\), \({F_{L}}\).
   The relevant angular variables are the angle \(\theta_{l}\) defined
   as the angle between the positive (negative) muon momentum and the
   direction opposite to the \(\Bdb\,(\Bd)\) in the dimuon reference
   frame and the angle \(\theta_{K}\) defined as the angle between the
   kaon momentum and the direction opposite to the \(\Bdb\,(\Bd)\) in
   the \(\Kstarzb\,(\Kstarz)\) rest frame.  The decay rate distribution
   of \(\Bdb\to{\Kstarzb}\mumu\) is described as a function of
   \(\theta_{l}\) and \(\theta_{K}\) and is measured in several
   \({q^{2}}\) (\(\equiv{m^{2}(\mumu)}\)) bins. The main results of the
   analysis, \({F_{L}}\) and \({A_{FB}}\) are extracted from unbinned
   extended maximum likelihood fits to three variables: \(m(\Bdb)\)
   and the two angular variables.  The results yielded by fits for every
   \({q^{2}}\) bin are compared to SM
   predictions~\cite{Kruger:1999xa}. Deviations from the SM predictions
   may indicate new phenomena.
\par  
   The CMS analysis~\cite{cms-conf-afb} uses the data sample of
   \(\IntL\approx5.2\,\ifb \) collected by several flavors of CMS dimuon
   trigger.  The CMS trigger acceptance criteria are
   \(\left|\eta(\mupm)\right|<2.2\), \(\pt(\mupm)>3,\,4,\,4.5,\,5\gevc\)
   (depending on trigger flavor) and \(\pt(\mumu)>6.9\gevc\). The CMS
   trigger fits \(\mumu\) pairs to a common point required to be
   displaced from a vertex of the origin of \(\propro\) interaction. The
   \(\Kstarzb\) candidates are reconstructed through their decay mode
   \(\Kstarzb\to\Km\pip\)
   and the \(\Bdb\ra{\Kstarzb{(892)}}(\ra{\Km\pip})\mumu\)
   is reconstructed by fitting the identified \(\mumu\) pair and the two
   hadron tracks each with \(\pt(h)>0.75\gevc\) to a common vertex.
\par  
   The ATLAS analysis~\cite{atlas-conf-afb} is based on
   \(\IntL\approx4.9\,\ifb\) of data collected by ATLAS single muon and
   dimuon triggers.  The main ATLAS triggers select di-muon events
   requiring both muons to have \(\pt(\mupm)>4\gevc\) or alternatively
   \(\pt(\mu_{1})>6\gevc\) and \(\pt(\mu_{2})>4\gevc\) with a rapidity
   range of \(\left|\eta(\mupm)\right|<\)2.4. In most ATLAS triggers no constraint
   on the di-muon invariant mass is applied. The \Kstarzb candidates are
   reconstructed from hadron tracks with \(\left|\eta(h)\right|<\)2.5 and
   \(\pt(h)>0.5\gevc\).
\par  
   The results are shown in Fig.~\ref{fig:fl-afb}.  No deviations from
   the SM predictions~\cite{Kruger:1999xa} are found by both
   experiments. There is a slight tension between ATLAS data and SM
   expectations for low \({q^{2}}\) bins.
  \begin{figure}[hbt]
  \begin{center}
       \subfigure{%
            %\label{fig:cms-prod-yns}
         \includegraphics[width=0.25\textwidth]{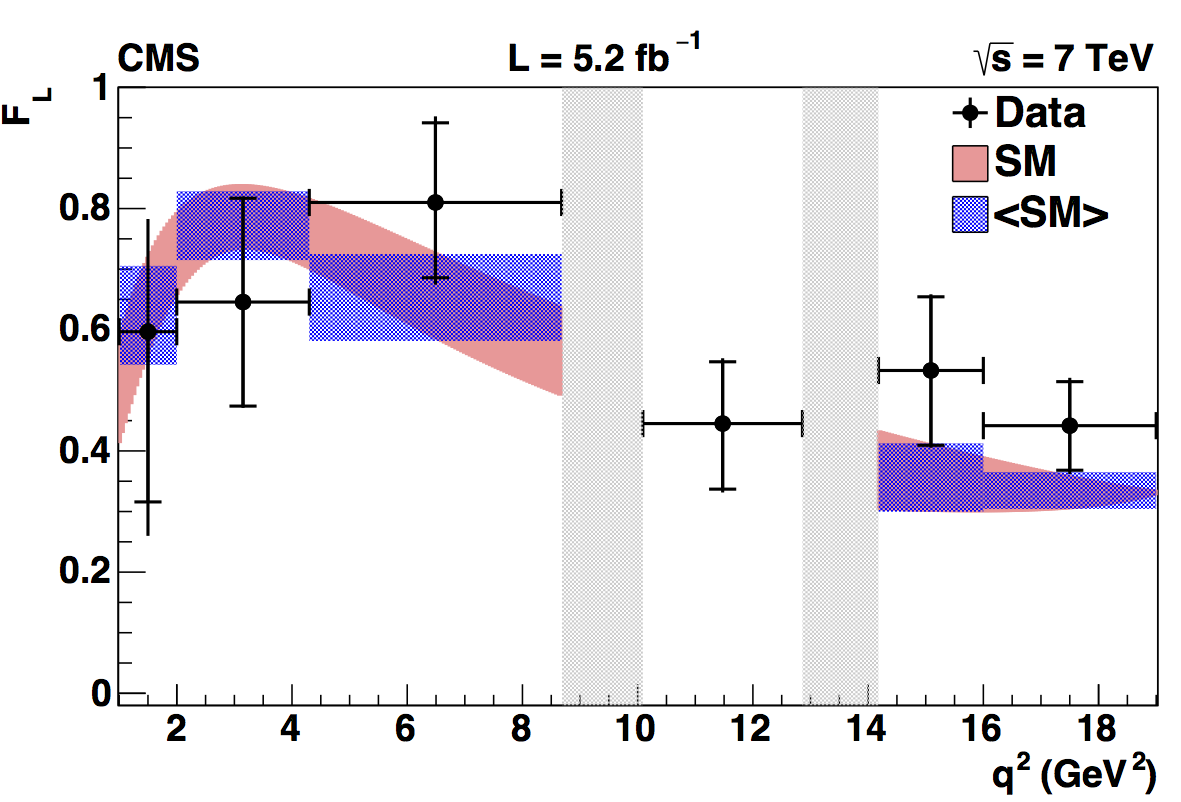}
       }%
       \subfigure{%
            %\label{fig:cms-prod-yns}
         \includegraphics[width=0.25\textwidth]{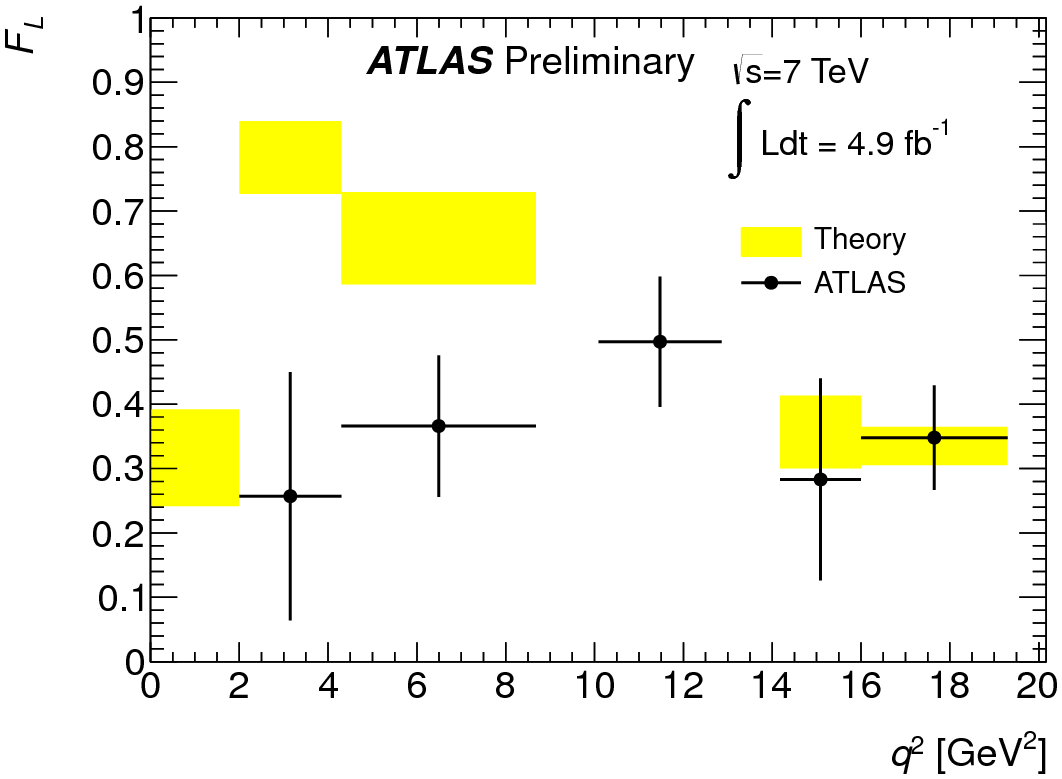}
       }\\%
       \subfigure{%
            %\label{fig:cms-prod-yns}
         \includegraphics[width=0.25\textwidth]{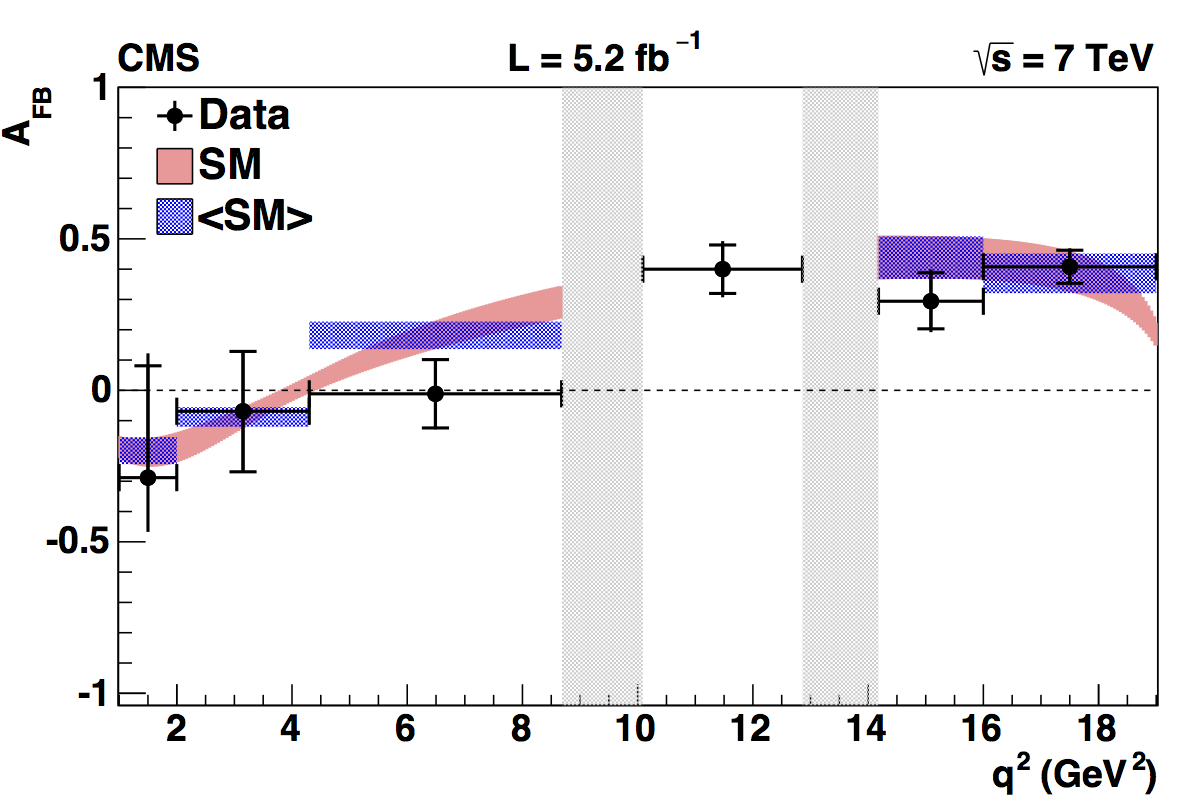}
        }%
       \subfigure{%
            %\label{fig:cms-prod-yns}
          \includegraphics[width=0.25\textwidth]{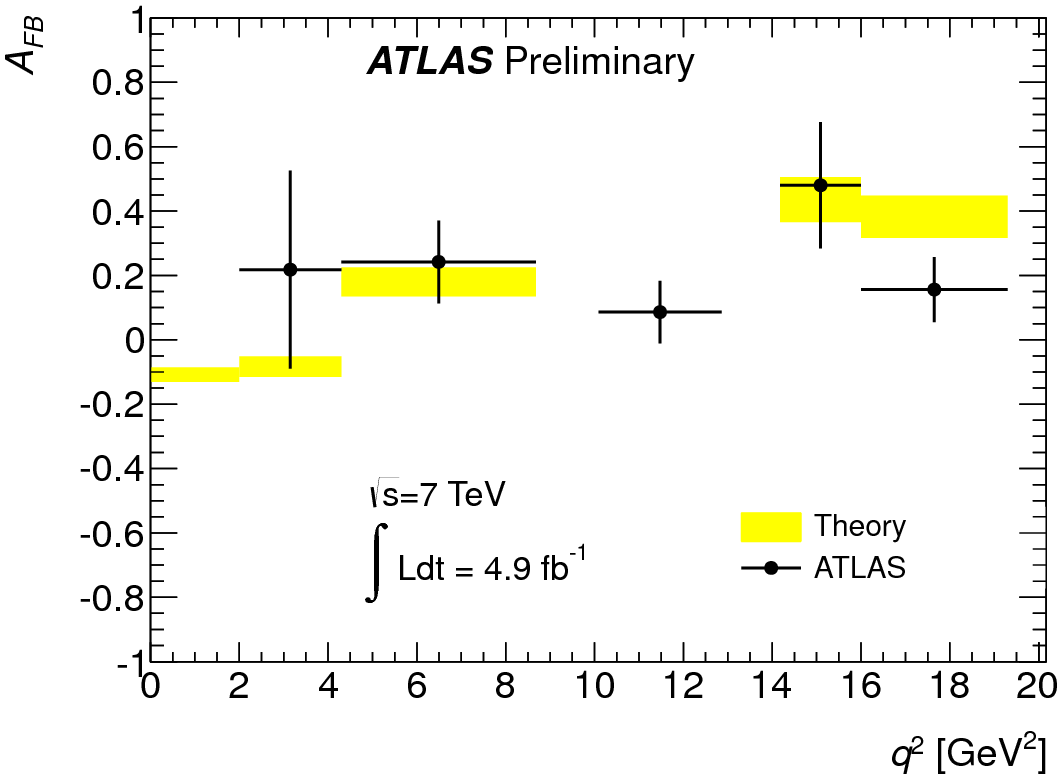}
       }%
  \end{center}
  \caption{ Results of the measurement of \({F_{L}}\) (upper plots) and
    \({A_{FB}}\) (bottom plots) versus the dimuon \({q^{2}}\).  The
    \({q^{2}}\) bins corresponding to \(\Jpsi \) and \(\psitwos\)
    resonances are not filled.  Good agreement with SM predictions~\cite{Kruger:1999xa}
    is found. }
  \label{fig:fl-afb}
  \end{figure}
\section{\(\Y1S,\,\Y2S,\,\Y3S\) cross section measurements by CMS}
\label{y123s}
  In this section we present the production of the lowest
  \(\Upsilon(nS)\) states in \(\propro \) collisions studied with the CMS
  detector. These bottomonium (\(\bbbar\)) states are produced promptly,
  contrary to charmonium (\(\ccbar\)) states originating partially from
  weak \b-decays.  Their dominant production mechanism is through the
  fragmentation of partons, \(\exegrat\), gluons, \(\glue\to\bbbar\) though some
  fraction of \(S\)-wave states result from the strong or radiative
  decays of higher, \(\exegrat\), \({P}\)-wave \(\bbbar\) states.  There are
  several bottomonium production models which predict different shapes
  of \(\pt\) production spectra at their high range in \(\propro \)
  collisions~\cite{Cho:1995vh}.  The experimental measurements from LHC
  provide access to high-\pt range of \(\Upsilon(nS)\) production
  spectra to make a useful comparison with the predictions.
\par  
  Earlier, the ATLAS Collaboration has published the production
  cross section measurements of all three \(\Y1S,\,\Y2S,\,\Y3S\)
  states~\cite{Aad:2012yna}. The analysis was based on
  \(\IntL\approx1.8\,\ifb\) collected by ATLAS dimuon triggers. The total
  production cross-section over \(p_{\rm T}^{\Upsilon}<70\gevc \) and
  \(\left|y^{\Upsilon}\right|<2.25\) and the \(\pt \) spectra in central
  rapidity \(\left|y^{\Upsilon}\right|<1.2\) and forward
  \(1.2<|y^{\Upsilon}|<2.25\) rapidity intervals have been
  measured~\cite{Aad:2012yna}.
\par  
  The latest analysis by the CMS Collaboration presented here is based on an
  unprescaled dimuon trigger involving the tracker and muon
  systems~\cite{cms-conf-yns}. The larger data sample of
  \(\IntL\approx4.9\,\ifb\) contains \(\Upsilon(nS)\ra\mumu\) candidates
  with vertex fitted and produced in a central area,
  \(\left|y^{\Upsilon}\right|<0.6\) with a broad \(\pt\) range of
  \(10-100\gevc\). The experimental spectra are found from fits of the
  invariant mass \(m(\mumu)\) distributions of the candidates
  reconstructed in several \pt bins of \(\left|y^{\Upsilon}\right|<0.6\)
  rapidity range.  The raw spectra are corrected by the trigger
  efficiency, by the acceptance of analysis criteria and normalized to
  the luminosity and \(\BR(\Upsilon(nS)\ra\mumu) \).
  The results are shown in Fig.~\ref{fig:prod-yns} for \(\Y1S\) and
  \(\Y2S\) states~\cite{cms-conf-yns}.  The \pt-spectra reveal at
  \(\pt\gsim{20}\gevc\) the change of an exponential shape to a
  power-law behavior.
  To emphasize the transition area the production
  ratio
  \({\frac{d\sigma}{d{p_{T}}}}\cdot\BR(\Y2S)/{\frac{d\sigma}{d{p_{T}}}}\cdot\BR(\Y1S)\)
  is presented at the bottom right plots both for CMS and
  ATLAS. Interestingly enough, the recently published ATLAS
  spectra~\cite{Aad:2012yna} do show the similar break-down of the shape
  around \(\pt\sim20\gevc \) as the right bottom plot of
  Fig.~\ref{fig:prod-yns} demonstrates. In a summary, new measurements
  of
  \({\frac{d\sigma}{d{p_{T}}}}\Big|_{|y|<0.6}\times{\BR\big(\Upsilon(nS)\to\mumu\big)}\)
  in a wide range of \(\pt\in(10-100)\gevc\) are made with CMS detector.
  The transition from nearly exponential cross section decrease with
  \(\pt \) to power-law behavior for all three \(\Upsilon(nS)\) is
  observed presenting a challenge to theoretical models.
\begin{figure}[hbt]
  \begin{center}
       \subfigure{%
            %\label{fig:cms-prod-yns}
         \includegraphics[width=0.25\textwidth]{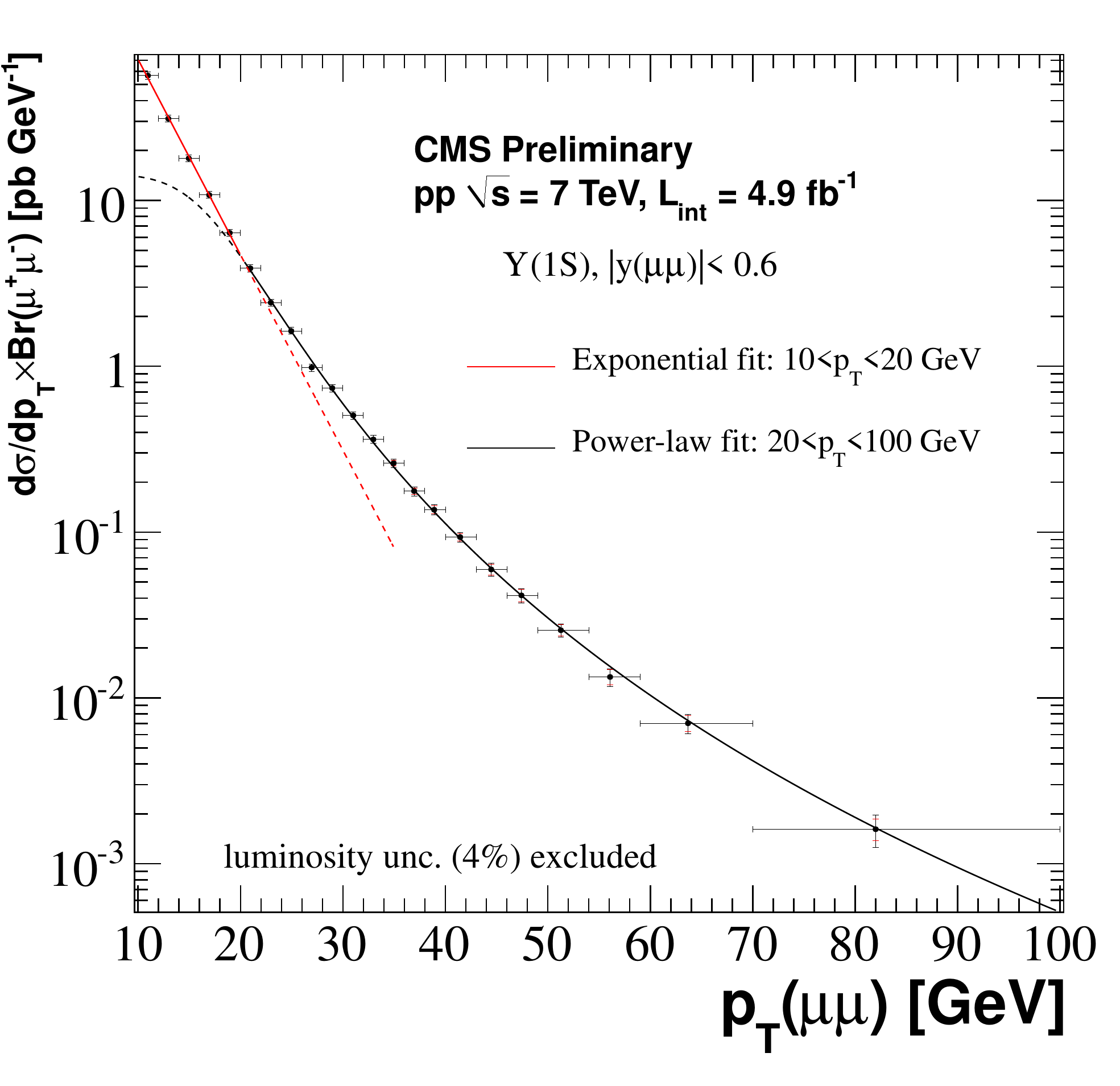}
       }%
       \subfigure{%
            %\label{fig:cms-prod-yns}
         \includegraphics[width=0.25\textwidth]{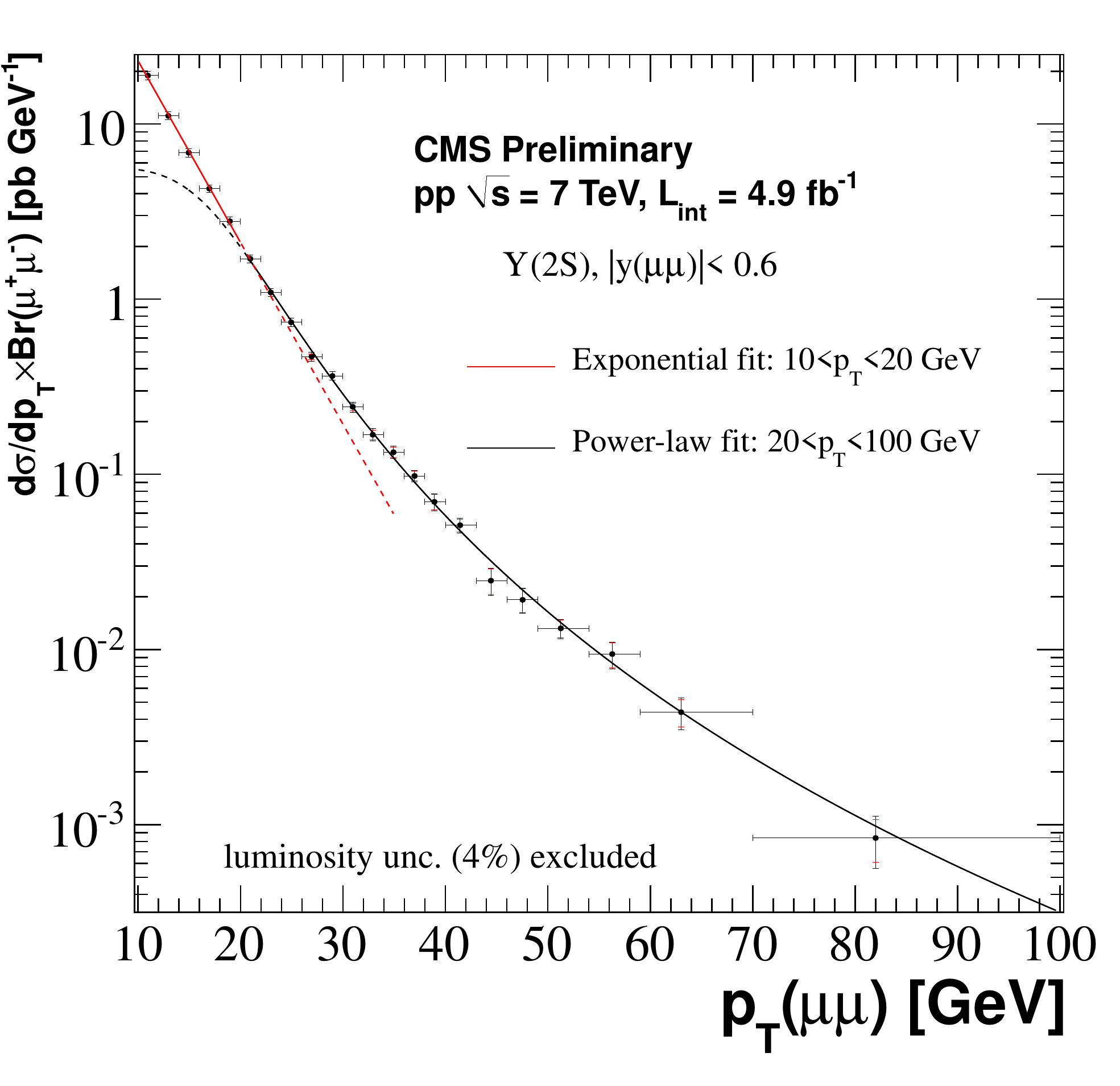}
       }\\%
       \subfigure{%
            %\label{fig:cms-prod-yns}
         \includegraphics[width=0.25\textwidth]{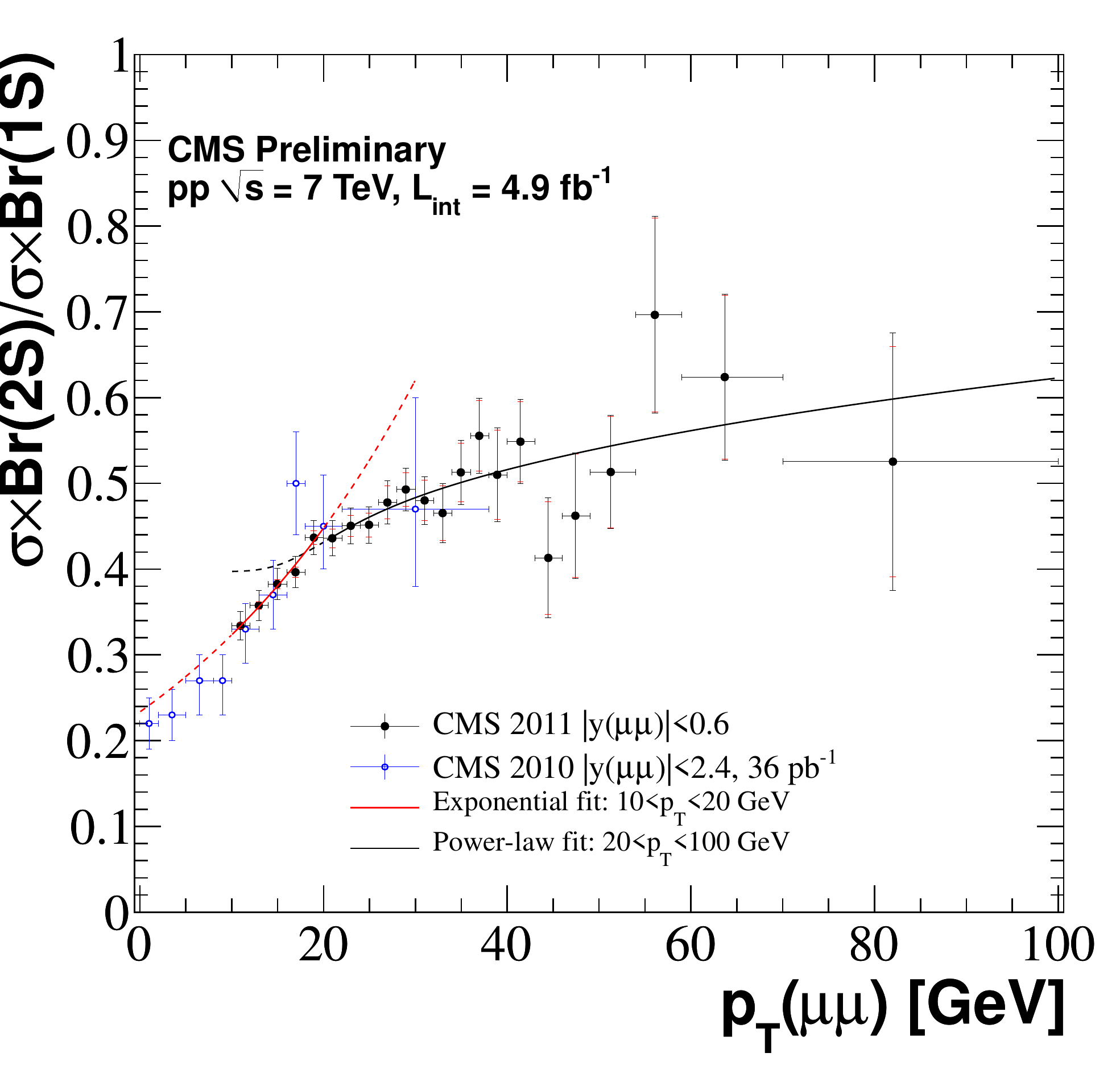}
        }%
       \subfigure{%
            %\label{fig:cms-prod-yns}
          \includegraphics[width=0.25\textwidth]{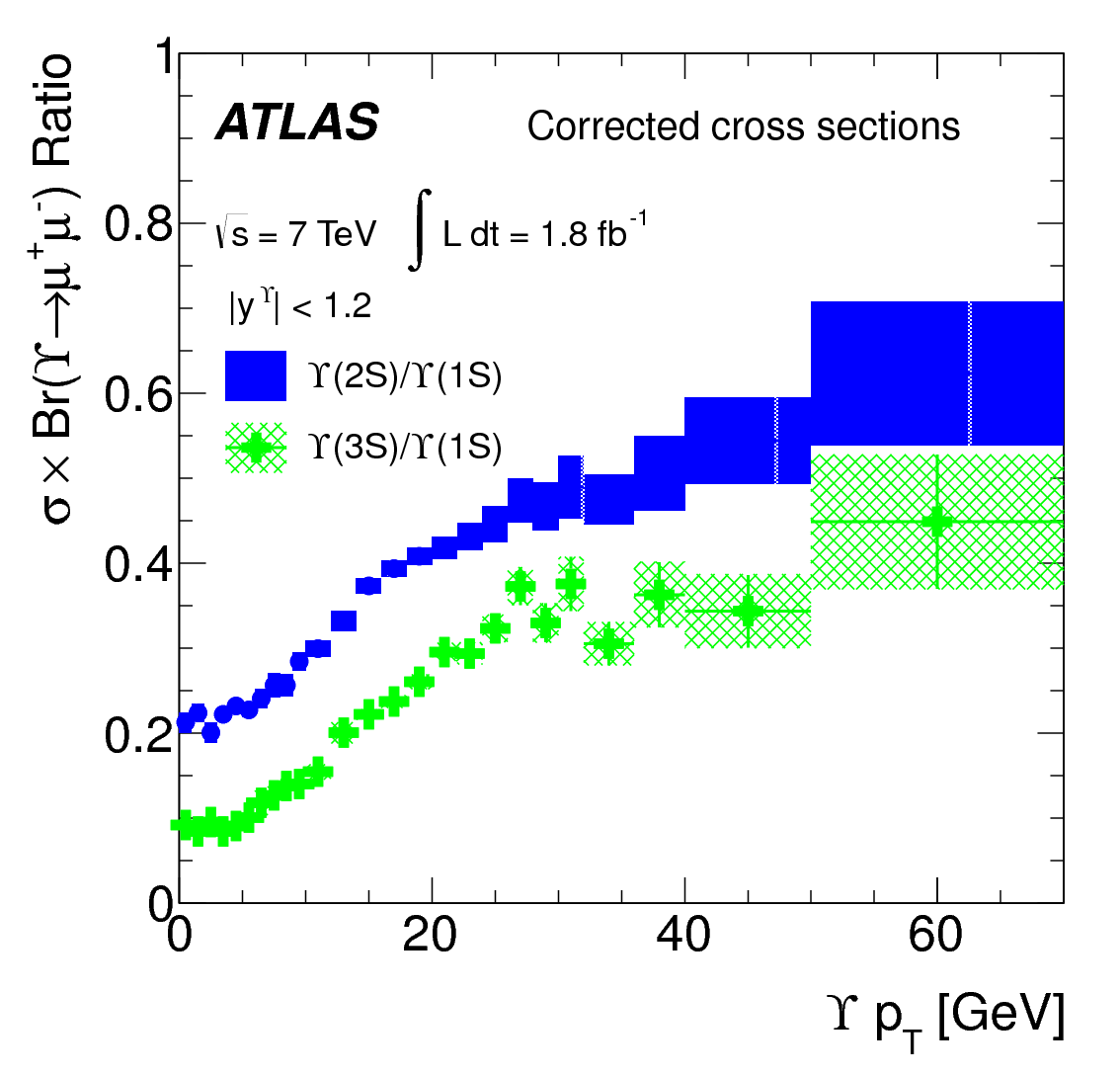}
       }%
  \end{center}
  \caption{ CMS results: \(\Y1S \) production cross section spectrum
            (upper left plot), \(\Y2S \) production spectrum cross section
            (upper right plot) and the ratio 
            \({\frac{d\sigma}{d{p_{T}}}}\cdot\BR(\Y2S)/{\frac{d\sigma}{d{p_{T}}}}\cdot\BR(\Y1S)\)
            of production spectra (bottom left plot).
            ATLAS result to compare: ATLAS
            \((\Y2S)/(\Y1S)\) 
            production ratio (bottom right plot). The same change in
            shape of the production ratio at high \(\pt \) is observed.}
  \label{fig:prod-yns}
\end{figure}
\section{\(\Bu\to\Jpsi\Kp\) cross section measurements by ATLAS}
\label{bpxsec}
  Measurements of the \b-hadron production cross section in \(\propro\)
  collisions at LHC provide further tests of QCD calculations for
  heavy-quark production at higher center-of-mass energies and in wider
  transverse momentum (\pt) and rapidity (\(y\)) ranges, thanks to the
  extended coverage and excellent performance of the LHC
  detectors. ATLAS and CMS are sensitive in the central rapidity region,
  so their measurements are complementary to open beauty measurements
  with LHCb detector.
\par  
  In this section we present the production cross section measurement
  for \(\Bu \) reconstructed in its fully exclusive decay mode to
  \(\Jpsi(\ra\mumu)\Kp \) using the ATLAS detector. The data for this
  analysis correspond to an integrated luminosity
  \(\IntL\approx2.4\,\ifb\) collected at \(\sqs=7\tev \) by a dimuon
  trigger, which requires the presence of at least two muon candidates
  of \(\pt(\mupm)>4.0\gevc\) and \(\left|\eta(\mupm)\right|<2.4\) each.  Offline,
  the events are required to contain at least one pair of reconstructed
  muons, and each pair is fitted using a vertexing algorithm. The
  corresponding \(\Jpsi\ra\mumu \) candidate with fitted common vertex
  is selected with the invariant mass, \(m(\mumu)\in(2.7,\,3.5)\gevcc \).
  The muon tracks of the selected \(\Jpsi\) are again fitted to a common
  vertex with an additional hadron track of \(\pt(h)>1\gevc\) and with
  the \(\Kpm\) mass assigned.  The three-track vertex fit is performed
  by constraining the muon tracks to the \(\Jpsi \) world average
  mass~\cite{Beringer:1900zz}.  The \(\Bu\) candidates with
  \(\pt(\Bu)>9\gevc\) and \(\left|{y(\Bu)}\right|< 2.3\) in the mass range
  \(m(\Bu)\in(5.040,\,5.800)\gevcc\) are kept for further analysis.
  The number of reconstructed \(\Bu\) mesons is obtained using a binned
  maximum likelihood fit to the invariant mass of the selected
  candidates per every \((\Delta{\pt},\,\Delta{y})\) bin.
  The cross section \(d^{2}\sigma(\propro\ra\Bu\,+X)/d{\pt}d{y}\) for
  four \(\Delta{y} \) and eight \(\Delta{\pt} \) intervals, covering the
  range of \(\left|{y}\right|< 2.25\) and \(\pt\in(9,\,120)\gevc \) are
  presented in Fig.~\ref{fig:prod-bp}~\cite{ATLAS:2013cia}.
  The measured differential cross section is compared with the
  {\sc QCD NLO} calculations.  The predictions are obtained using
  {\sc Powheg+Pythia} and {\sc MC$@$NLO+Herwig} and are quoted with an
  uncertainty from renormalization and factorization scales and \b-quark
  mass of the order of \((20-40)\% \). Within these uncertainties,
  {\sc Powheg+Pythia} predictions are in agreement both in absolute scale
  and in the shape with the measured \(\pt\) and \(y\) double
  differential distributions. At low \(\left|{y}\right|\), 
  {\sc MC$@$NLO+Herwig} predicts lower production cross section and a
  softer \(\pt \) spectrum than the one observed in data, which becomes
  harder for \(\left|{y}\right|>1.0\).  An {\sc FONLL} calculation
  with \({f_b}\rightarrow\Bu=(40.1\pm1.3)\%\) to fix the overall scale,
%  
  %assuming a hadronization fraction of \({f_b}\ra\Bu\) of
  %\((40.1\pm1.3)\%\) to fix the overall scale, 
%  
  is in a good agreement with the measured spectra, in particular at 
  \(\pt<30\gevc\) range.
\begin{figure}[hbt]
  \begin{center}
       \subfigure{%
            %\label{fig:cms-prod-yns}
            \includegraphics[width=0.25\textwidth]
            {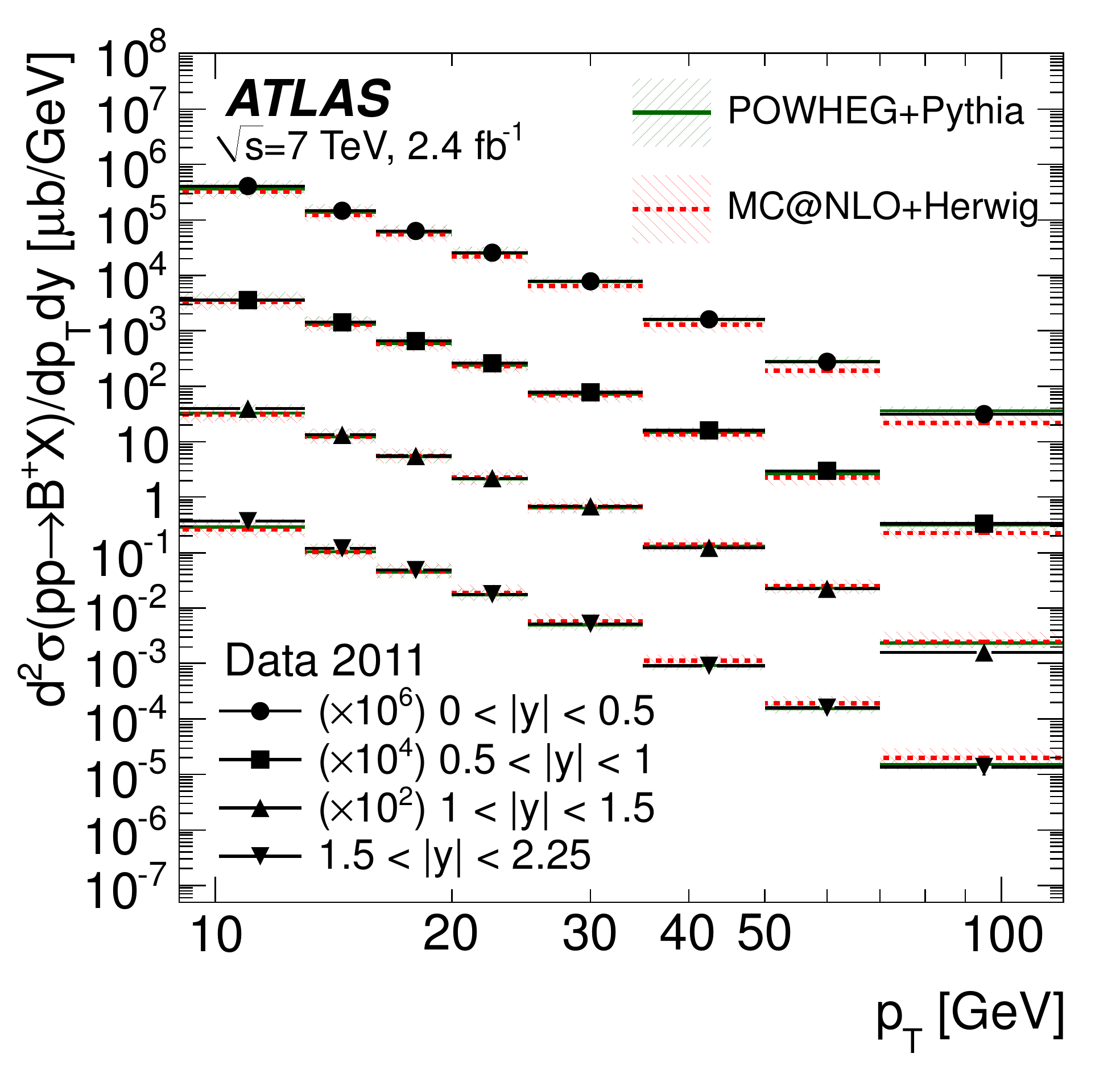}
       }%
       \subfigure{%
           %\label{fig:atlas-prod-yns}
           \includegraphics[width=0.25\textwidth]
           {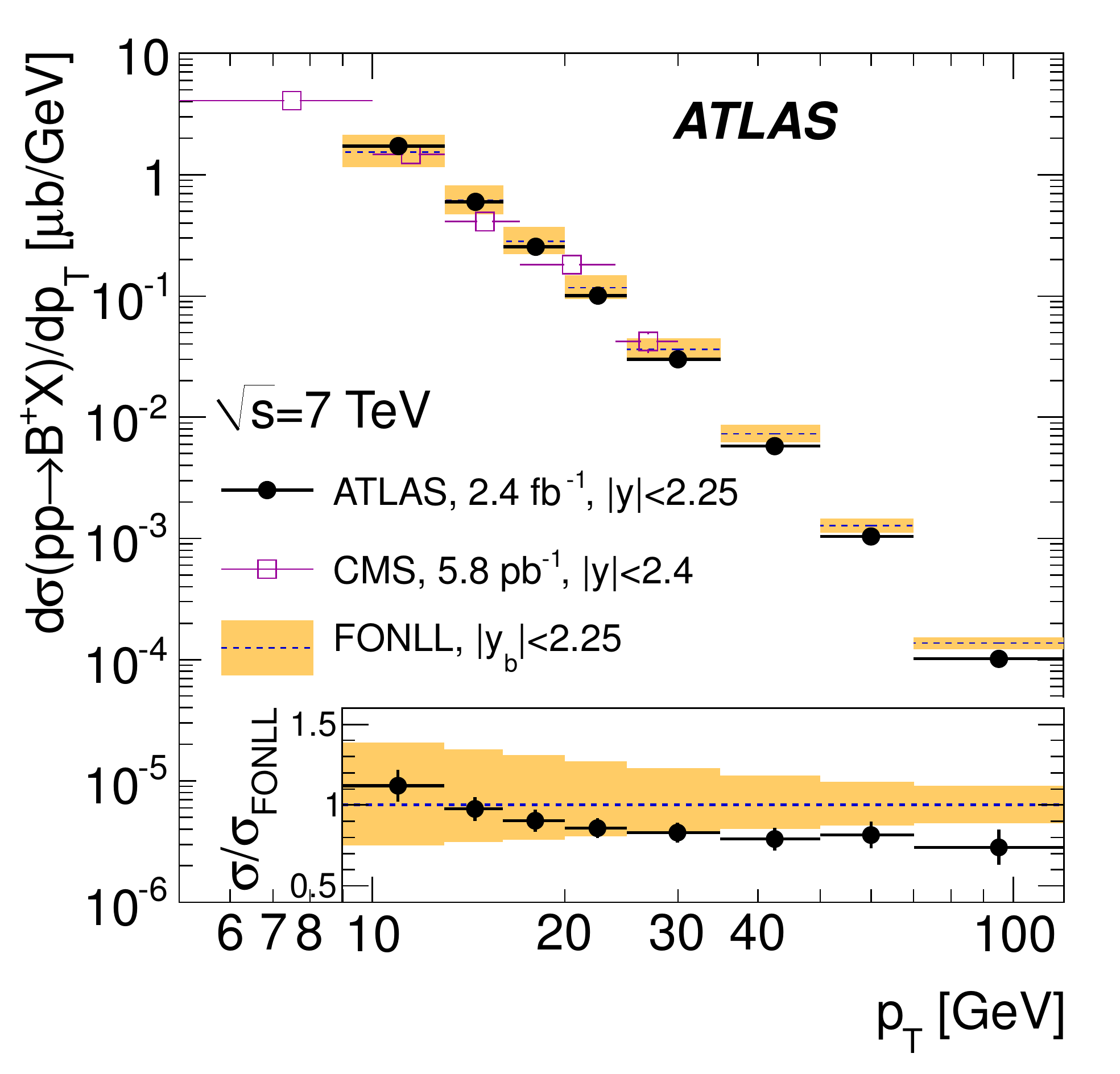}
       }%
  \end{center}
  \caption{ Double-differential cross section of \Bu production as a
            function of \pt for different rapidity ranges: (left plot) The data
            points are compared to {\sc NLO} predictions from {\sc Powheg} and
            {\sc MC$@$NLO}; 
            (right plot) the data points are compared with 
            {\sc  FONLL} predictions (see also inset) and 
            the CMS results as well.
  }
  \label{fig:prod-bp}
\end{figure}
%%
%% 
%
%%
%
%% References with BibTeX database:
%\nocite{*}
%\bibliographystyle{elsarticle-num}
%\bibliography{martin}

%% Authors are advised to use a BibTeX database file for their reference list.
%% The provided style file elsarticle-num.bst formats references in the required Procedia style

%% For references without a BibTeX database:

\end{document}

%%
%% End of file `nuphbp-template.tex'. 